\documentclass[twocolumn,showpacs,nofootinbib]{revtex4}
\usepackage[utf8]{inputenc}
\usepackage{amssymb,eucal,amsthm,epsf,graphics,float}
\usepackage[english]{babel}
\def\e{{\mathop{\,\rm e}}}

\def\sgn{{\mathop{\rm sgn}}}
\def\sh{{\mathop{\rm sh}}}
\def\vf{v_{\!f}}

\begin{document}

\title{Symmetries of the quasi-1d Bechgaard salts superconducting state in an applied magnetic field}


\author{N. Belmechri}
\affiliation{Laboratoire de Physique des Solides,
UMR 8502 CNRS – Université Paris-Sud 11}
\author{G. Abramovici}
\affiliation{Laboratoire de Physique des Solides,
UMR 8502 CNRS – Université Paris-Sud 11}
\author{M. Héritier}
\affiliation{Laboratoire de Physique des Solides,
UMR 8502 CNRS – Université Paris-Sud 11}


\begin{abstract}

We investigate the possible superconducting pairing symmetries which would account for the latest experimental results on the Bechgaard salts. Using a renormalization group (R.G) technique we calculate the couplings of the singlet and triplet interactions in the quasi 1D Hamiltonian. In the presence of interchain interactions, the singlet and triplet couplings are of the same order of magnitude. The renormalized couplings are then used in an RPA calculation to evaluate the critical fields of superconductivity for the two dominant order parameter symmetries in the R.G flow, i.e, the $d_{x^2 - y^2}$ and $f_y$ symmetries. It is shown that, for the standard values of the anisotropy ratio of the Fermi surface, the critical fields of the singlet symmetry are strongly reduced. A reentrant superconductivity is however still present for the triplet state. 
\end{abstract}


\maketitle

\paragraph*{ Introduction}

Since the discovery twenty eight years ago of superconductivity in the organic compounds of the $(TMTSF)_2X$, family ( X= $PF_6$, $ClO_4$, ) \cite{mazaud,ishiguro}, these compounds have been intensively studied both theoretically and experimentally. Many original phenomena has been discovered and studied. However, at the time being even the symmetry of the superconducting state is not yet known with certainty. The simplest question of which pairing symmetry characterizes the order parameter is still a matter of debate. Many apparently contradicting experimental results have attempted to address this question. Many have suggested the spin-pairing of electrons to be of the singlet type, while others suggested the spin-triplet. Our purpose, in this paper, is to study which theoretical model for the superconducting state is able to account for all the experimental data.

At zero magnetic field, the effect of non magnetic impurities \cite{joo} has been shown to reduce the critical temperature, in agreement with an unconventional symmetry, and with the presence of nodes of the gap function on the Fermi surface \cite{behnia}.

Measurements of the critical field in the $a$ and $b'$ \cite{leehope,oh,lee1,lee2} directions performed on both $PF_6$ and $ClO_4$ compounds have reported a superconducting state surviving up to fields as high as 9 and 5 $T$ respectively. In $a$ and $b'$ directions, the maximum critical fields reached, are two times higher than the Pauli paramagnetic limiting field. At high magnetic fields $\sim 4 T$, an increase in the critical temperature in increasing magnetic field has been previously observed, which was ascribed to a reentrant superconducting phase \cite{leehope}.

Lee \textit{et al.} measured the electron spin contribution to the total $^{77}Se$ Knight-shift in the $PF_6$ compound for fields parallel to the $a$ and $b'$ directions respectively \cite{lee3}. In both cases it was found that the resonant frequency shift remains unchanged when the system passes through the superconducting transition, whereas it is expected to decrease and ultimately vanish if the pairing were of the singlet type. Later, Shinagawa \textit{et al.} \cite{shinagawa} performed analogous measurements on $ClO_4$ compound for a field strength of 1.38 $T$, and drew the same conclusions. However, very recently, new Knight-shift measurements performed by Shinagawa \textit{et al.} on the $ClO_4$ at lower field values (0.9 $T$) in the $a$ and $b'$ directions, showed a decreasing spin susceptibility upon cooling through the superconducting transition, while it remained unchanged for a field of 4 $T$. 

On the theoretical level, there have been three main ideas proposed to interpret the experimental results. The first idea was that the pairing symmetry of the order parameter is singlet at all magnetic fields. The high field regime is only marked by a first order transition \cite{dupuis1,dupuis2,dupuis3} from a homogeneous superconducting state to a LOFF state \cite{shimahara4,shimahara5,shimahara,Maki1} where the Cooper pair has a finite total momentum in order to compensate for the Zeeman effect. In the second interpretation, the superconducting state is assumed to be in the triplet \cite{lebed2,lebed3,Maki2} Equal Spin Pairing (ESP) state at all magnetic fields. Using this symmetry, with an anisotropic $\vec{d}$-vector order parameter, Lebed \textit{et al.} \cite{lebed2} could explain the high field Knight-shift results. It has also been suggested \cite{shimahara2,melo,poilblanc,noomen} that in an increasing magnetic field the symmetry of the order parameter could change from singlet to triplet so that the superconductivity is no longer paramagnetically limited. In our previous work \cite{noomen}, we could calculate the transition line between a $d$-singlet and a $p$-triplet superconducting state using the free energy criteria. There, we considered the Zeeman effect as the only limiting effect on superconductivity, based on the fact that at the transition field the confinement of the electron orbital motion to the conduction layers reduces considerably the orbital effect of the magnetic field on superconductivity. This field induced dimensional crossover has been discovered by Lebed in his pioneering work of ref. \cite{lebed5}, which is essential to the physics discussed here.

In the present paper we study a $d_{x^2-y^2}$ singlet and an $f_y$ triplet state at the vicinity of the critical field, when the field is parallel to the $b'$ direction. First, we will start our calculation by evaluating the renormalized triplet and singlet couplings using a renormalization group approach. Then, we will use the renormalized couplings in the calculation of the critical fields, for both symmetries of the order parameter.

\paragraph*{Theoretical model}

In order to explain the 1D behavior of the Bechgaard salts at high temperatures, it is well known that the coupling between Cooper channel (electron-electron pairing) and Peierls one (electron-hole pairing) has to be taken into account. This is possible by means of a renormalization group (RG) approach \cite{solyom}, used to calculate the effect of the high T 1D fluctuations on the different couplings. When temperature is decreased, the system undergoes a dimensional cross-over at $T=T^\ast\sim300K$ where its 3D character becomes predominant \cite{solyom}. The Cooper and Peierls channels are then decoupled, and the RG flow may be turned into a standard RPA calculation. When a magnetic field is present, a three cut-off RG approach \cite{heritier0,haddad} is more appropriate. In the following we use this RG method in order to determine the renormalized singlet and triplet couplings.

At high temperature, we use the standard g-ology model, with hamiltonian
$H=H_{\rm kin}+H_{\rm int}$, where kinetic and bare interaction parts write
\begin{eqnarray}
H_{\rm kin}&=&\sum_{\bf k\;\sigma\;\alpha}\xi_{\bf k}^{\alpha}
a^{\dag\alpha}_{\bf k\sigma}a^{\phantom{\dag}\alpha}_{\bf k\sigma}\qquad\hbox{with}\nonumber\\
\label{eq1}
\!\!\!\!\!\!\!\!\!\!\!\!\!\!\xi_{\bf k}\,=\,
\vf(|k_x|-k_f)&-&2t_b\cos(k_y b)\;-\;2t_c\cos(k_z c)
\end{eqnarray}

\begin{eqnarray*}
H_{\rm int}&=&
\sum_{\bf k k' q\atop\alpha\;\alpha'}
{g_s({\bf k,k'})\over2}
a_{\bf{q\over2}-k\uparrow}^{\alpha\dag}
a_{\bf{q\over2}+k\downarrow}^{-\alpha\dag}
a_{\bf{q\over2}+k'\uparrow}^{\alpha'}
a_{\bf{q\over2}-k'\downarrow}^{-\alpha'}\\
&+&
\sum_{\bf k k' q\atop\alpha\;\sigma}
g_t({\bf k,k'})
a_{\bf{q\over2}-k\;\sigma}^{\alpha\dag}
a_{\bf{q\over2}+k\;\sigma}^{\alpha\dag-}
a_{\bf{q\over2}+k'\,\sigma}^{\alpha}
a_{\bf{q\over2}-k'\,\sigma}^{-\alpha}\\
&+&
\sum_{\bf k k' q\atop\alpha\;\sigma\;\sigma'}\!\!\!
{g_t\over2}({\bf k,k'})
a_{\bf{q\over2}-k\;\sigma}^{\alpha\dag}
a_{\bf{q\over2}+k\;-\sigma}^{\alpha\dag-}
a_{\bf{q\over2}+k'\,\sigma'}^{\alpha}
a_{\bf{q\over2}-k'\;-\sigma'}^{-\alpha}\\
&-&\sum_{\bf k k' q\atop\alpha\;\alpha'}
g_{bb}
a_{\bf{q\over2}-k\uparrow}^{\alpha\dag}
a_{\bf{q\over2}+k\downarrow}^{-\alpha\dag}
a_{\bf{q\over2}+k'\uparrow}^{\alpha'}
a_{\bf{q\over2}-k'\downarrow}^{-\alpha'}cs\\
&+&
\sum_{\bf k k' q\atop\alpha\;\sigma}
2g_{bb}
a_{\bf{q\over2}-k\;\sigma}^{\alpha\dag}
a_{\bf{q\over2}+k\;\sigma}^{\alpha\dag-}
a_{\bf{q\over2}+k'\,\sigma}^{\alpha}
a_{\bf{q\over2}-k'\,\sigma}^{-\alpha}cs\\
&+&
\sum_{\bf k k' q\atop\alpha\;\sigma\;\sigma'}\!\!\!
g_{bb}
a_{\bf{q\over2}-k\;\sigma}^{\alpha\dag}
a_{\bf{q\over2}+k\;-\sigma}^{\alpha\dag-}
a_{\bf{q\over2}+k'\,\sigma'}^{\alpha}
a_{\bf{q\over2}-k'\;-\sigma'}^{-\alpha}cs
\end{eqnarray*}
where $cs=\cos(\bold(k_y +k'_y).b)$.

The kinetic Hamiltonian is linearized around the Fermi points with a Fermi velocity $\vf$. $a$, $b$ and $c$ are the lattice parameters in
the $x$, $y$ and $z$ directions, where for simplicity we assumed an orthorombic structure. $\alpha=(+/-)$ stands for left and right moving electrons respectively. $\sigma$ is the electron spin.

In the interaction Hamiltonian, the first three terms are related to intrachain
interactions, while the last three ones are related to interchain interactions.
In these interchain terms, $g_s$ and $g_t$ are replaces by $g_{bb}$,
the backward interchain scattering, which operates as a modulation of
the intrachain ones.

The three cut-off RG method involves three characterictic energies, $\mu_B H\ll k_B T^\ast\ll\Lambda_0$ ($\Lambda_0$ is the energy band-width). For $T>T^\ast$, the system is effectively 1D, and the energy scale of the magnetic field as well as that of $t_c$ are irrelevant.

We use the One Particle Irreductible (OPI) scheme \cite{Metzner,Honerkamp,nickel,Abramovici} to
recalculate the scattering couplings in a one-loop expansion. The main idea of
this RG method is to sum up high energy momenta in the action in order to get an
effective action in terms of low energy momenta. This is done by reducing
the energy scale $\Lambda=\Lambda_0\e^{-\ell}$, during the flow ($\ell$ is the
flow parameter and the flow begins with $\Lambda=\Lambda_0$). For each
$\Lambda$, we get effective couplings $g_i(\Lambda)$ ($i=s,t$) while 
$g_i(\Lambda_0)$ are the bare couplings.
Using the OPI scheme, we write the RG equations relating the ${dg_i\over dt}(\Lambda)$ to the $g_i(\Lambda)$. The lengthy RG equations will not be listed here, for details see Ref. \cite{nickel,Abramovici}. In our numerical calculation of the renormalized couplings, there are three independant adjustable parameters, $t_b$, the bare values of $g_i$ (where we chose bare
couplings $g_s\sim3g_t$ which does not affect the final point in the flow) and the inter-chain backward scattering coupling $g_{bb}$. We have used the following values $g_s/(\pi\vf)=1.04$, $g_t/(\pi\vf)\sim0.4$ and $g_{bb}/(\pi\vf)=0.038$ which are compatible with known experimental data. At $\Lambda=\Lambda^\ast \simeq t_b$, which fairly corresponds to $T\sim T^\ast$, we get the renormalized couplings : ${g_s}^*= 0.361$ and ${g_t}^*=0.312$. For $T\le T^\ast$, the couplings between Cooper and Peierls channels become negligible, so that the RG equations are equivalent to those obtained in RPA. Because of the decoupling of the RG equations for $T\le T^\ast$, the values obtained for the renormalized couplings at $\Lambda^*$ remain unchanged for $T\le T^\ast$.

\begin{figure}[h]
$\!\!\!\!$\rotatebox{-90}{\scalebox{0.35}{\includegraphics{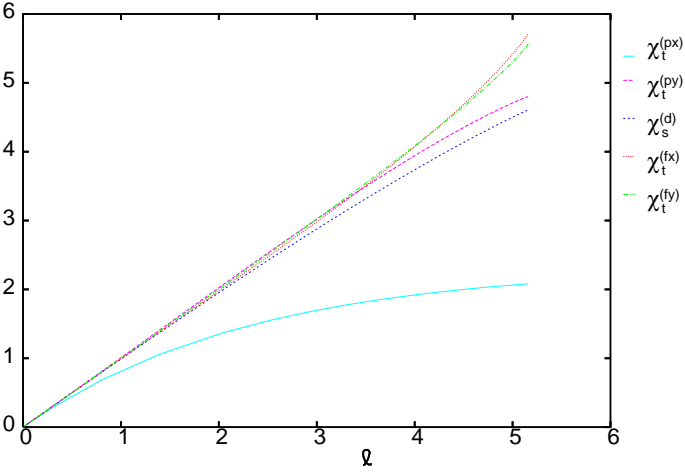}}}
\caption{Flow of the superconducting susceptibilities $\chi^{(d)}_s$ (singlet of
symmetry $d$), $\chi^{(p_x)}_t$ (triplet of symmetry $p_x$),  $\chi^{(p_y)}_t$
(triplet of symmetry $p_y$),  $\chi^{(f_x)}_t$ (triplet of symmetry $f_x$)  and
$\chi^{(f_y)}_t$ (triplet of symmetry $f_y$), with bare parameters
$g_s/(\pi\vf)=-1.04$, $g_t/(\pi\vf)=-0.4$, $g_{bb}/(\pi\vf)=0.038$ and
$2t_b/\Lambda_0=0.133$.}
\label{susc_RG}
\end{figure}

If we continue the RG flow up to $\Lambda_c$, where some couplings begin to diverge, we get the following singlet and triplet response functions (see Fig.~\ref{susc_RG}), where one can observe the competition between triplet $f$-pairing susceptibilities and singlet $d$-pairing ones. In fact, by increasing the CDW fluctuations, the inter-chain backward scattering enhanced the $f$-triplet superconductivity, resulting in renormalized triplet couplings of the same order of magnitude as those of the singlet ones.

Renormalized couplings, ${g_s}^*$ and ${g_t}^*$, will be used in the following low temperature effective mean field hamiltonian.


\begin{eqnarray}
H  &=& H_{\rm kin}+H_{s,t}\nonumber\\
\label{eq2}
H_s&=&  \sum_{\bold k \bold q \alpha} \Delta_s (\bold k,\bold q ) a_{({\bold q} / 2)-\bold k, \uparrow }^{\alpha ~ \dag } a_{({\bold q} / 2)+\bold k \downarrow }^{-\alpha ~ \dag } + h.c
\\
H_t&=&  \sum_{\bold k \alpha} \Delta_t (\bold k ) \{ a_{\bold k, \uparrow }^{-\alpha ~ \dag } a_{-\bold k, \uparrow }^{\alpha ~ \dag } + a_{-\bold k, \downarrow }^{\alpha ~ \dag } a_{\bold k, \downarrow }^{-\alpha ~ \dag }  \} + h.c \nonumber
\end{eqnarray}

where singlet and triplet order parameters are

\begin{eqnarray}
\label{eqsing}
\Delta_s (\bold k,\bold q)&=&\Delta_{ \uparrow \downarrow} (\bold k, \bold q)\\\nonumber
&=&  \sum_{\bold k',\alpha} {g_s}^\ast (\bold k, \bold k') ~ \langle ~ a_{{{({\bold q} / {2})}+{\bold k'}}, \downarrow }^{-\alpha}~ a_{{{({\bold q} / 2)}-{\bold k'}}, \uparrow }^{\alpha} ~ \rangle \nonumber
\end{eqnarray}

\begin{eqnarray}
\label{eqtrip}
\Delta_t (\bold k)&=&\Delta_{\uparrow \uparrow} (\bold k)=-\Delta_{ \downarrow \downarrow} (\bold k)  \\\nonumber
&=&  \sum_{ \bold k',\alpha } {g_t}^\ast (\bold k, \bold k') ~ \langle ~ a_{{{-\bold k'}}, \uparrow }^{\alpha}~ a_{{{\bold k'}}, \uparrow }^{-\alpha}\rangle  \nonumber
\end{eqnarray}

$\bold k$ is related to the internal degrees of freedom of the pairing state and gives the symmetry of the wave function, while for the singlet order parameter $\bold q$ is related to the motion of the center of mass of the Cooper pair and gives the spatial variations of the gap function. $\bold q$ is therefore the LOFF wave vector.

The $d_{x^2-y^2}$ and $f_y$ symmetries, and the corresponding interaction channels, are given by

\begin{eqnarray}
\label{eqpp}
\Delta_s (\bold k, \bold q)&=&\Delta_s (\bold q) ~ \cos(k_{\perp})\\
{g_s}^\ast(\bold k,\bold k')&=&~{g_s}^\ast ~\cos(k_{\perp}) ~\cos(k'_{\perp}) \\ \nonumber
\\
\quad\Delta_t (\bold k)&=&\Delta_t ~ \sin(2k_{\perp})\\
\label{eq.}
\quad ~{g_s}^\ast(\bold k,\bold k')&=&{g_s}^\ast ~\sin(2 k_{\perp}) ~ \sin(2 k'_{\perp}) 
\end{eqnarray}

We rewrite the linearized energy dispersion Eq.(\ref{eq1}) as

\begin{equation}
\label{eq3}
\xi_{\bold k,\sigma}^{\alpha}  =  v_{f\sigma}^{x}(k_y) \left[\alpha k_x - k_{f\sigma}^{x}(ky)\right]-2 t_c~\cos(c k_z)
\end{equation}
where Fermi velocity and Fermi wave number now depend on the transverse position on the Fermi surface $k_y$ as

\begin{eqnarray*}
k_{f\sigma}^{x}(k_y) & = & k_f + {{2 t_b} \over \vf} ~ \cos(b k_y) + {{\beta t_b} \over \vf} ~ \cos^2 (b k_y) \\
& + & {{{\sigma \mu_b H} \over {\vf} }} \left[ 1 + \beta ~\cos(b~ k_y)\right] \\
v_{f\sigma}^{x}(k_y) & = & 2 a t_a \sin\left[a k_{f\sigma}^{x}(k_y)\right]
\end{eqnarray*}

$k_{f\sigma}^{x}(ky)$ is the $x$ component's module of the Fermi wave vector of the spin $\sigma$ electrons, and $v_{f\sigma}^{x}(ky)$ is the corresponding Fermi velocity. $\beta = \sqrt{2}~ {t_b /t_a}$. $\mu_b$ is the electron magnetic moment. $a$, $b$ and $c$ are the lattice parameters. 

Unlike the above RG calculation, in the low temperature regime, i.e below the critical temperature for superconductivity $T_c (0) \sim 1 K$, the magnetic field is relevant and should be considered in the mean field hamiltonian Eq.(\ref{eq2}). Its Zeeman effect is already included through the spin dependance of the energy spectrum. The orbital effect will be fully taken into account through a Peierls substitution, which is justified in our case. Moreover, Eq.(\ref{eq3})  takes into account the real quasi-1d form of the Fermi surface as was first noticed by Lebed \cite{lebed6}, in contrast to previous calculations where a flat 1D Fermi surface was used \cite{melo,dupuis1,dupuis2,dupuis3,lebed5}.

In ref. \cite{noomen} we have studied the hamiltonian Eq.(\ref{eq2}) with the Zeeman splitting as the only effect of the magnetic field. There, we neglected the transverse transfer integral $t_c$ compared to $t_a$ and $t_b$, and therefore neglected the orbital effect of the magnetic field. By calculating the gap equations, we could study the $d_{x^2-y^2}$ and $p_y$ symmetries of the superconducting order parameter. $p$, $d$ and $f$ symmetries are believed to be the most probable candidates for the symmetry of the superconducting state in the Bechgaard salts \cite{ishiguro,joo,fuseya,kuroki}. The gap equations for the $f$ order parameter and for the $p$ one \cite{lebed6} are identical. Using the free energy criteria, we have been able to calculate a first order phase transition line between the $d$-singlet and the $p$-triplet superconducting states. However, in our calculation of the phase transition we did not take into account the possible occurence of a LOFF state which should succeed to the homogeneous singlet state in increasing magnetic fields. Moreover, from the above RG calculations it appears that $f$-wave symmetry is to be considered rather than the $p$ one.

In this paper, we will establish the gap equations for a $d_{x^2-y^2}$ and a $f_y$ order parameters in a magnetic field in the $b'$ direction. 

The electron Green's function in the $b'$ direction is given in the gauge $\bold A=(0,0,-H x)$ \cite{lebed10,heritier} by 

\begin{eqnarray}
&&g_{\sigma}^{0\alpha}(x,x';{\bold k}_{\perp} ; i \omega_n)={{-i ~\sgn(\omega_n)} \over {\vf}}  \\
&\times& \exp\left[{{i\alpha (x-x')}\over{\vf}} (i\omega_n-\epsilon_f) \right] \nonumber \\
&\times& \exp\;{i\alpha}\left[ \lambda_z ~ \cos\{c k_z + G {(x+x')/ 2}\}~ \sin\{G{{x-x'}/ 2}\} \right] \nonumber 
\end{eqnarray}

where $\alpha\omega_n (x-x')>0$, $G=eHc$, e the electron charge, and $\lambda_c = 4 t_c/(\vf G)$ is the semi classical amplitude of the electron motion in the interplane direction $c$ \cite{lebed5}. 

Using the equation of motion for the above Green's function we can write the linearized gap equations for $(T-T_c(H)) \ll T_c(H)$ as follows

\begin{eqnarray}
\label{eqd}
\Delta_{q_x}(x)&=&\displaystyle
{{g_s}^\ast \over 2}\int^{^\infty}_{\atop{\atop\!\!\!\!\!\!\!\!\!\! \vert x'-x \vert > {d_c}}} \textrm{d}x' {{2 \pi T} \over { \vf ~\sh\left[{2 \pi T (x'-x) }\over \vf \right]} }\\
\nonumber\\
&\times& \left( J_0 \left[\tilde{\beta} (x-x')\right]-J_2 \left[\tilde{\beta} (x-x')\right] \right)  \nonumber\\
\nonumber\\
& \times& J_0 \left[~ 2 \lambda_z \sin(G (x-x')/ 2) \sin( G (x+x')/ 2)~ \right]\nonumber\\
\nonumber\\
& \times& \cos({{2 \mu_b H}\over{\vf}}(x-x'))\Delta_{q_x}(x')\nonumber
\end{eqnarray}

for singlet order parameter, and

\begin{eqnarray}
\label{eqfy}
1&=&\displaystyle
{{g_t}^\ast \over 2}\int^{^\infty}_{\atop{\atop\!\!\!\!\!\! {d_c}}} \textrm{d}x {{ 2 \pi T} \over { \vf ~\sh\left[{2 \pi T x }\over \vf \right]} }\\
\nonumber\\
& \times& J_0 \left[~ 2 \lambda_z \sin^2 (G x/ 2) )~ \right]\nonumber
\end{eqnarray}


for the triplet order parameter.

${d_c}$ is a cut-off distance. $\tilde{\beta}=(2 \mu_b H  / \vf) \beta$ . ${g_{s,t}}^\ast$ are the dimensionless interaction constants determined in the RG calculation. The singlet superconducting gap function was assumed to have the form $\Delta_{\bold q}(\bold k, \bold r)= \Delta_{\bold q}( \bold r)\cos(k_{\perp})$. And we anticipated the fact that the maximum critical field corresponds to the LOFF wave vector perpendicular to the field direction, so that $\Delta_{\bold q}( \bold r ) = \Delta_{q_x}(x)=\Delta ~ \cos( q_x x) $. It should be noted that the $f_y$-gap equation (\ref{eqfy}) is identical to the $p_y$-gap equation calculated previously \cite{lebed6}. The only difference resides in the values of the $p$ and $f$ couplings.



\paragraph*{Results and discussion}

The integral in the $f$-wave gap equation (\ref{eqfy}) diverges for $T \rightarrow 0$, indicating that there is no finite value for the critical field. This is at the origin of the reentrance phase of superconductivity at high magnetic fields for the triplet case. On the contrary, the integral in the $d$-wave gap equation (\ref{eqd}) converges at $T=0$, and therefore the zero temperature critical field for the singlet phase exists and is finite, this is the paramagnetic limit of the LOFF state, i.e $H_c^{LOFF}$. 

Let us first revisit the case without orbital effect, where $t_c=0$. Taking the limit $T \rightarrow 0$ in Eq.(\ref{eqd}) we find

\begin{eqnarray}
\ln\left[\displaystyle{H_c^{LOFF}\over H^\ast}\right] &=&
\int^{\infty}_0 \textrm{d} x ~ \left[(J_{0} \left[\beta x\right]-J_{2} \left[\beta
x\right])\cos(k x)-1\right]\nonumber\\
\nonumber\\
&\times&(\cos(x) / x)
\end{eqnarray}

with $\left( \ln[2 \mu_b H^{*} d_c / \vf]={2 \over g_s}-\gamma \right)$ where
 $\gamma$ is the Euler constant. $\\$

\begin{figure}[!h]
$\!\!$\rotatebox{-90}{\scalebox{0.35}{\includegraphics{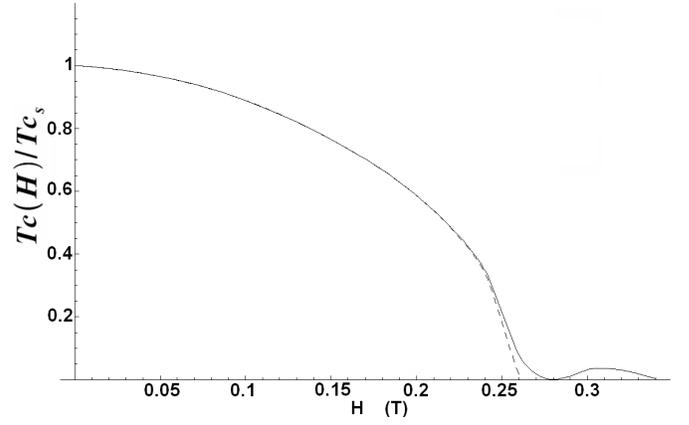}}}
\caption{ $b'$-direction critical temperature vs magnetic field of the $d_{x^2-y^2}$ symmetry for $t_a = 3000 K$ and $t_c = 5K$. The dashed line shows the critical field for the homogeneous state, while the solid line shows the LOFF-state critical field after maximization with respect to the LOFF wave vector. $T_{cs}\sim1.1 K$ is the critical temperature corresponding to the singlet coupling constant $g_s ^\ast$}
\label{d1}
\end{figure}

\begin{figure}[!h]
$\!\!\!\!\!\!\!\!$\rotatebox{-90}{\scalebox{0.38}{\includegraphics{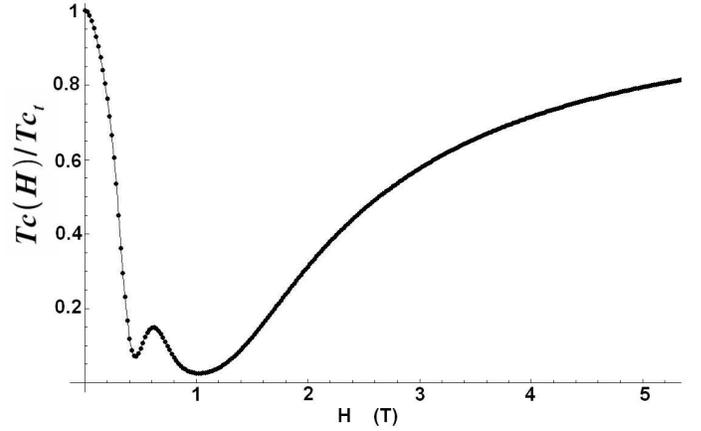}}}
\caption{$f_{y}$ $b'$-direction critical temperature vs magnetic field of the $f_y$ symmetry for $t_a = 3000 K$ and $t_c = 5K$. $T_{ct}\sim0.5 k$ is the critical temperature corresponding to the triplet coupling constant $g_t^\ast$}
\label{f1}
\end{figure}

For our numerical calculations we have used the standard in-plane transfer integral values, $t_a : t_b \simeq 3000 : 300  K$ \cite{ishiguro}. After maximization with respect to the LOFF wave vector k, we find $H_{c}^{LOFF}(t_c=0)=5.3~ T$ for a LOFF wave vector $q_x = {{2 \mu_b H}\over \vf} (1-\beta)$. 

\begin{figure}
$\!\!\!\!\!\!$\rotatebox{-90}{\scalebox{0.365}{\includegraphics{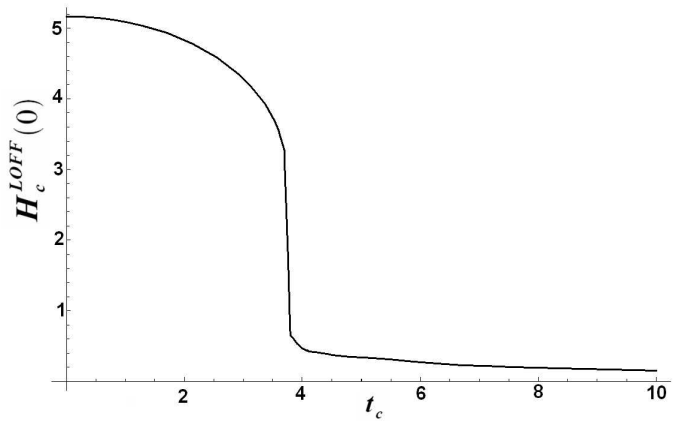}}}
\caption{ zero temperature critical field dependance on $t_c$. The $t_c\sim3.8 K$ value, at which there is a sudden drop of $H_c ^{LOFF} (0)$, corresponds to a semiclassical amplitude $\lambda_z $ reaching exactly the interplane distance $c$, which corresponds to the deconfinement of electrons out of the $a-b$ planes.}
\label{ft}
\end{figure}

The expected effect of a finite $t_c$ is to further reduce the value of $H_{c}^{LOFF}$. In the calculation of the critical fields, the values used in the litterature \cite{miyasaki,shinagawa, grant, heritier0} for $t_c$ range from 2 K to 17 K, and those used for the anisotropy ratio $t_a/t_c$ range from $\sim 200$ to $\sim 700$. However, the maximum LOFF critical field, $H_{c}^{LOFF}(T=0)$, depends strongly on the value of the $t_a/t_c$ ratio Fig.~\ref{ft}. Therefore, in the subsequent numerical calculations, we will limit ourselves to values for $t_c$ in the range $5-10 K$, which are the closest values to experimental and theoretical determinations for $t_c$ \cite{grant,heritier0}. In that case $H_c^{LOFF}$ at $T=0$ is reduced to the very low value of $0.34 T$. Then, a LOFF state could not be stable and the system is in a triplet state at high fields, because of the reentrance phenomena. However, $H_c^{LOFF}$ is so sensitive to $t_c$, which is not precisely known, that $t_c\lesssim3.5 K$ would be enough to get $H_c^{LOFF}\sim2.5 T$. We cannot exclude this possibility, which would ensure the existence of a LOFF state and of a triplet state at high fields.

\begin{figure}[-H]
$\!\!\!\!\!\!\!\!$\rotatebox{-90}{\scalebox{0.355}{\includegraphics{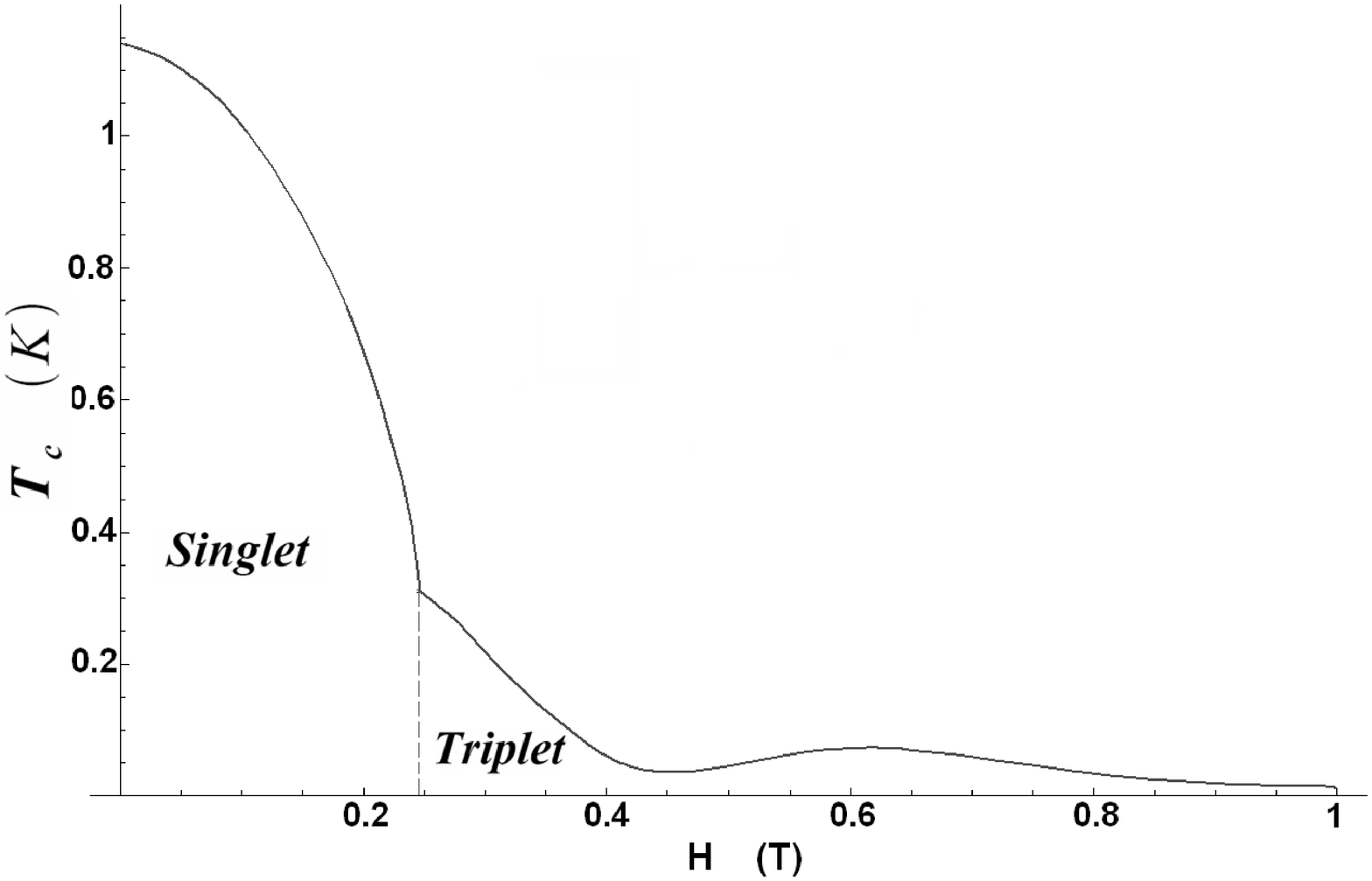}}}
\caption{ Schematic $H-T$ phase diagram calculated from equations (\ref{eqd}) and (\ref{eqfy}).}
\label{fdiag}
\end{figure}

By numerically solving the gap equations (\ref{eqd}) and (\ref{eqfy}), we have been able to calculate $T_c (H)$ for the homogeneous and LOFF $d_{x^2 - y^2}$ phases as well as for the $f_y$ phase Fig. \ref{d1} an \ref{f1}. The most striking result is that, for the standard anisotropy ratio $t_a / t_c \simeq 600$, the maximum critical field of the LOFF phase is strongly reduced by the anisotropic orbital effects, as can be seen on Fig. \ref{d1}. Therefore, the value of the anisotropy ratio is crucial to the determination of the critical field of the LOFF phase. In order to reproduce the critical fields measured experimentally we have to use very high anisotropy ratios which do not correspond to the commonly known values. For the standard values indeed, $\lesssim 700 $, the LOFF phase is limited to very low magnetic fields, Fig. \ref{ft}. This could qualitatively explain the recent NMR measurements by Shinagawa \textit{et al.}, where at a field of $0.9 T$ the Knight-shift in the $ClO_4$ is consistent with singlet symmetry, while at slightly higher fields, $\sim 1.4 T$, the Knight shift is that of a triplet symmetry. Then, our results support the hypothesis of a phase transition from singlet to triplet superconductivity that takes place at relatively low fields, as can be seen on Fig. \ref{fdiag}. It should be noted that a first order phase transition is also expected from homogeneous to LOFF superconductivity \cite{dupuis1,dupuis2}. However, the proximity of the two transitions, Fig. \ref{d1}, could make it difficult to discriminate between them.

For the $f_y$-triplet phase, we find the same reentrance phenomenon, Fig. \ref{f1}, as that previously found for the $p_y$ symmetry \cite{lebed6,dupuis1,dupuis2}. At high magnetic fields, the triplet pairing of the Cooper electrons and the absence of Zeeman effect are at the origin of the re-stabilization of superconductivity. This phenomenon has already been observed in the $ClO_4$ compound by Lee \textit{et al.} \cite{leehope}. Besides the reentrance and the oscillations of $T_c$ at low temperatures, our calculations show that, at intermediate magnetic fields, the critical temperature is strongly reduced, Fig. \ref{f1}. This is clearly an orbital effect, and may be improved if we take into account the magnetic field dependance of the coupling constants, which may not be negligible in our case.

\paragraph*{Conclusion}

We have used an RG calculation to determine the singlet and triplet coupling constants which we have used in the calculation of the critical fields. We show that for a not too high anisotropy ratio $t_a / t_c$, the critical fields for a d-wave LOFF state are strongly reduced, while a reentrant superconductivity is expected for the f-wave triplet state. From these results we conclude that a transition from the singlet LOFF state to the triplet state at low magnetic fields seems to be the most reliable interpretation of the recent experimental results.

More experimental work is still needed though. In particular, more measurements of the Knight-shift as a function of magnetic field at low temperatures are necessary. Specific heat measurements will also be needed in order to observe the phase transition.

We would like to thank J. Friedel, J. C. Nickel, S. Charfi-Kaddour and S. Haddad for stimulating discussions and valuable help.


\end{document}